\documentclass[12pt,aps,floats,showpacs,amssymb,tightenlines]{revtex4}

\usepackage{amsmath}
\usepackage{amsfonts}
\usepackage{amscd}
\usepackage{epsfig}
\usepackage{amssymb}
\usepackage{tabularx}
\def\a{\alpha}
\def\b{\beta}

\def\d{\delta}
\def\l{\lambda}
\def\L{\Lambda}

\def\p{\partial}
\def\k{\kappa}
\def\s{\sigma}
\def\r{\rho}

\begin{document}

\title{Plane fronted gravitational waves in Lovelock-Yang-Mills theory}
\author{Reinaldo J. Gleiser and Gustavo Dotti}
\affiliation{Facultad de Matem\'atica, Astronom\'{\i}a y F\'{\i}sica,
Universidad Nacional de C\'ordoba, Ciudad Universitaria,
(5000) C\'ordoba, Argentina}
\email{gdotti@fis.uncor.edu}

\begin{abstract}
We obtain plane fronted gravitational waves (PFGWs) 
in arbitrary dimension in 
Lovelock gravity, to any order in the Riemann tensor. We exhibit pure
gravity as well as Lovelock-Yang-Mills PFGWs.
Lovelock-Maxwell and  $pp$ waves arise 
as particular cases. The electrovac  solutions
trivially satisfy  the Lovelock-Born-Infeld field equations.
The peculiarities that arise in degenerate Lovelock theories are also analyzed.
\end{abstract}

\maketitle

\noindent
\section{Introduction}

The classical field equation for the space-time metric in string
theory is the condition of conformal invariance of a two-dimensional
$\s$ model. It involves higher order terms in the curvature, which 
are expected to play a significant role in regions 
around singularities. As was shown by Lovelock 
in the early seventies \cite{Lovelock}, the possible corrections to Einstein
gravity are quite limited, since the only symmetric, 
divergence free tensor than can be constructed out of the metric and its
first two derivatives in a $d-$dimensional space-time is
\begin{equation} \label{lovelockt}
{\cal{G}}_b{}^a = \sum_{p=0}^{[(d-1)/2]} \a_p {G_{(p)}}_b{}^a, 
\end{equation}
where the $\a_p$'s are arbitrary constants and 
${G_{(p)}}_b{}^a$ is a tensor of order $p$ in the curvature, given by,
\begin{equation} \label{gp}
{G_{(p)}}_b{}^a \equiv  \d_{[b}{}^a 
R_{i_1 i_2}{}^{i_1 i_2} R_{i_3 i_4}{}^{i_3 i_4} \cdots 
R_{i_{2p-1} i_{2p}]}{}^{i_{2p-1} i_{2p}}.
\end{equation}
In (\ref{gp}) an implicit sum over repeated indices 
is understood after antisymmetrization (which includes a $1/(2p+1)!$ normalization factor).
${G_{(0)}}_{ab}, {G_{(1)}}_{ab}$ and ${G_{(2)}}_{ab}$ 
are respectively proportional to 
 the  space-time metric $g_{ab}$, 
the Einstein tensor $R_{ab} -\frac{1}{2} R g_{ab}$, 
and  the Gauss-Bonnet tensor,
\begin{multline} \label{gb}
{G_{(2)}}_b{}^a \propto R_{cb}{}^{de}  R_{de}{}^{ca}  -2 R_d{}^c R_{cb}{}^{da}
-2 R_b{}^c R_c{}^a + R  R_b{}^a
 -\frac{1}{4} \d^a_b \left(
R_{cd}{}^{ef}R_{ef}{}^{cd} - 4 R_c{}^d R_d{}^c + R^2 \right)
\end{multline}
The field equations are 
\begin{equation} \label{lovelock}
{\cal{G}}_b{}^a = 8 \pi T_b{}^a,
\end{equation}
where $T_{ab}$ is the stress-energy tensor.
Einstein gravity (with a cosmological constant $\a_o$) is 
recovered if we  set $\a_p=0$ for  $p > 1$. The 
 $p>1$ terms appear naturally as higher order corrections in
 string theory  \cite{str1}. Stringy corrections 
higher than quadratic in the Riemann tensor are considered 
in \cite{str2}, and 
the role of the quartic Lovelock term in M-theory is 
discussed in \cite{m}. A BRST approach to Lovelock gravity 
can be found in \cite{jeno}, where it is also shown that adding 
terms involving covariant derivatives of the Riemann tensor to
the Lovelock action 
does not change the linearized equations  around a Minkowskian background.\\
In this paper we find all  PFGW solutions 
of Lovelock gravity coupled to a (possibly trivial) 
source free Yang-Mills field with gauge group $G$. As particular cases 
we get electrovac ($G=U(1)$) and
 $pp$ waves. The latter  are  of current interest  in string theory 
because it is possible to obtain the string spectrum 
of a string moving in a $pp-$wave background (see,  e.g.,  \cite{malda} 
and references therein). Our electrovac solutions trivially satisfy
the Lovelock-Born-Infeld field equations.\\
The PFGW equations for Lovelock-Yang-Mills
theory
are worked out  in Section \ref{pfgw} of the paper and solved  in
Section~\ref{zc}  for the case 
of flat wave fronts, and in  Section~\ref{nzc} 
for wave fronts with a  non zero curvature. As far as we know, this 
is the first calculation of PFGWs in Lovelock gravity to any order. 
Yet, 
some previously known results arise as specific limits of ours, 
mainly,  the higher dimensional Einstein gravity PFGWs given in \cite{r}.
There is also some intersection   with our work
in \cite{joro}, where a
restricted class of plane wave solutions 
of Einstein equations 
is shown to satisfy (\ref{lovelock}) simply because 
all $p>1$ terms in (\ref{lovelock}) are trivial. The
 solutions we exhibit in this
work have ${G_{(p)}}_b{}^a \neq 0$ for all $p$.
A -by no means exhaustive- list of further related work
includes the gravitons in \cite{gg1},  the Einstein-Yang-Mills solutions 
in \cite{guven,gg2}, the 
Lovelock-Born-Infeld black holes  constructed in \cite{g,cm}, 
and the higher dimensional $pp$ waves studied in, e.g.,  \cite{pph}.

\section{PFGWs in Lovelock-Yang-Mills theory} \label{pfgw}
As defined in \cite{orr}, a PFGW is an  $n+2$ dimensional
 space-time with a congruence of null geodesics which is 
shear, expansion and twist free. The associated null vector field $k^a$ 
is orthogonal to $n-$dimensional space-like surfaces of constant 
curvature, and there is a -possibly trivial-
Yang-Mills field for which these surfaces are the wave fronts, and
$k^a$ 
the wave vector. As was shown in \cite{gdp,orr} and generalized
to $n>2$ in \cite{r}, such a space-time admits local 
coordinates where the line element reads 
\begin{equation} \label{geo}
ds^2 = -2 d\s \left( S d\s + d \r \right) \left( \frac{Q}{P} \right) ^2 +
\frac{1}{P^2} \sum_{i=1}^{n} (dz^i) ^2.
\end{equation}
Here $k^a = \p/ \p \r$, the wave fronts are the surfaces of constant $\s$, and 
\begin{eqnarray}\label{p}
P(z) &=& 1 + \frac{\l}{4} \sum_{i=1}^{n} (z^i) ^2 \\ \label{q}
Q(\s,z) &=& \left( 1 - \frac{\l}{4} \sum_{i=1}^{n} (z^i) ^2 \right) \a(\s)
+ \sum_{i=1}^{n} z^i \b_i(\s) \\ \label{s}
S(\s,\r,z) &=& -\frac{\r ^2}{2} \left[ \l \a(\s)^2 + \sum_{i=1}^{n} (\b_i(\s))^2 
\right]+ \r \frac{\p_{\s} Q}{Q}+ \frac{P^{N/2}}{2 Q} H(\s,z).
\end{eqnarray}
The wave
fronts are of  constant curvature, namely, the Riemann tensor corresponding
to the n-dimensional metric $ds^2_{(wf)} =
 {P^{-2}} \sum_{i=1}^{n} (dz^i) ^2$ satisfies,
 \begin{equation} \label{riemannwf}
R_{(wf)}{}_{ij}{}^{k \ell} = \lambda (\d_i{}^k \d_j{}^{\ell} - \d_j{}^k \d_i{}^{\ell}),
\end{equation}
and the scalar curvature is,
\begin{equation}
R_{(wf)} = n(n-1)\l,
\end{equation}
Therefore, $\l=0$ corresponds to flat wave fronts. 
The space-time (\ref{geo}) is a generalization to arbitrary dimensions 
of Kundt's class \cite{k1,k2}. The coordinate transformations 
preserving the form (\ref{geo}) were studied in \cite{k2}, 
where it was shown that we can either set $\a=1$ or $\a=0$, and that the sign 
of 
\begin{equation}
\k :=  \l \a(\s)^2 + \sum_{i=1}^{n} (\b_i(\s))^2 
\end{equation}
is invariant. The metrics (\ref{geo}) were classified in \cite{k2}
according to the signs of $\l$ and $\k$ (see also Section V in \cite{orr}).
For $\l=\k=0$, $\a(\s)=1$, 
we get a particular case of the PFGW
that corresponds
to a $pp-$wave:
\begin{equation} \label{ppgeo}
ds^2 = -2 d\s \left( H(\s,z) d\s + d \r \right) + 
\sum_{i=1}^{n} (dz^i) ^2.
\end{equation}
The null vector $k^a$ is covariantly constant in this case.\\
In what follows, 
latin indices from the middle of the alphabet run from one to $n$ and are 
raised and lowered using the Euclidean metric $g_{ij} = \d_{ij}$ and its
inverse. Indices from the beginning of the alphabet take the values 
$\r,\s$ and $i$. 
The  Riemann tensor of the metric (\ref{geo}) is 
\begin{equation} \label{riemann1}
R_{ab}{}^{cd} = \lambda (\d_a{}^c \d_b{}^d - \d_b{}^c \d_a{}^d) + K_{ab}{}^{cd},
\end{equation}
where the only non-zero components of $K_{ab}{}^{cd}$ are those 
trivially related by symmetry to 
\begin{equation} \label{riemann2}
K_{\s j}{}^{\r i} = P  \d_j{}^i \left(\p^m P \right) 
\left( \p_m S \right) - P^2 \; \p^i \p_j S 
- \left( \frac{P^2}{Q} \right) \left[ \left(\p^i Q \right) 
\left( \p_j S \right) + \left(\p^i S \right) 
\left( \p_j Q \right) \right].
\end{equation}
From (\ref{riemann1}) and (\ref{riemann2})
\begin{equation} \label{ricci1}
R_a{}^b = (n+1)\, \l \, \d_a{}^b + K_a{}^b
\end{equation}
Here  $K_a{}^b = K_{ac}{}^{bc}$, its only nonzero component being
\begin{equation} \label{ricci2}
K_{\s}{}^{\r} = n P  \left(\p^m P \right) 
\left( \p_m S \right) - P^2 \; \p^m \p_m S 
- 2 \left( \frac{P^2}{Q} \right)  \left(\p^m Q \right) 
\left( \p_m S \right),
\end{equation}
which, after using (\ref{p}),(\ref{q}) and (\ref{s}) reduces to
\begin{equation} \label{kk}
K_{\s}{}^{\r} = - \frac{P^{n/2+2}}{2 Q} \left[ \p^k \p_k H + \frac{n(n+2)
\lambda H}{4 P^2} \right]
\end{equation}
Finally, the Ricci scalar and Einstein tensor are 
\begin{eqnarray} \label{ricci3}
R &=& (n+1)(n+2) \, \l. \\ \label{G1}
{G}_b{}^{a} &=& -\frac{n(n+1)}{2} \l \d_b{}^a +K_b{}^a = 
- \frac{1}{4} {G_{(1)}}_b{}^a.
\end{eqnarray}
In view of the antisymmetrization (\ref{gp}) and the fact that 
the only nonzero components of $K_{ab}{}^{cd}$ are (\ref{riemann2}),
there are no terms in ${G_{(p)}}_b{}^a$ higher than linear in $K_{ab}{}^{cd}$.
${G_{(p)}}_{b}{}^a$
contains a $\l^p$ term proportional to $\d_b{}^a$ and 
a $\l^{p-1}$ term proportional to $K_b{}^a$: 
\begin{equation}
{G_{(p)}}_b{}^a = u(p,n) \l^p \d_b{}^a + v(p,n) \l^{p-1} K_b{}^a.
\end{equation}
After some combinatorics we get
\begin{equation}
u(p,n) = \frac{2^p \, (n+1)!}{(n+1-2p)!}, \hspace{1cm}
v(p,n) = \frac{- 2^{p+1} \, p\, (n-1)!}{(n+1-2p)!}
\end{equation}
Lovelock's tensor is 
\begin{equation} \label{egb2}
{\cal{G}}_b{}^a =  F_1(\l) \;\d_b{}^a + F_2(\l)\; K_b{}^a
\end{equation}
where
\begin{eqnarray} \label{egb2_2}
F_1(\l) &=&  \sum_{p=0}^{[(d-1)/2]} \a_p u(p,n) \l^p \\
F_2(\l) &=& \sum_{p=1}^{[(d-1)/2]} \a_p v(p,n) \l^{p-1}. \nonumber
\end{eqnarray}
Then, if we impose that the wave front curvature $\l$
be related to the theory constants $\alpha_p$ through
\begin{equation} \label{egbm1}
 \sum_{p=0}^{[(d-1)/2]} \a_p u(p,n) \l^p  = 0
\end{equation}
the Lovelock equations (\ref{lovelock})  take the form
\begin{equation} \label{set}
T_{ba} = \frac{1}{8 \pi} 
\left( \sum_{p=1}^{[(d-1)/2]} \a_p v(p,n) \l^{p-1} \right) \;K_{ba} 
\end{equation}
The right hand side in (\ref{set})
 can be interpreted as the stress energy tensor of a gauge field with 
gauge group $G$ and potential 
 ${\cal A} = -\phi^B(\sigma,z)d\s \, T_B$, with $T_B$, ($B=1,...,d_G$) a basis
of $Lie(G)$. The field strength for this potential reduces to 
${\cal F} = d {\cal A} = 
 \p_k \phi^B\; d\s \wedge d z^k \, T_B$. This field is required to
be source free, so that there are no further  contributions 
to $T_{ab}$. The source free condition reads
\begin{equation} \label{egbm2}
 \p_k ( P^{2-n} \p^k \phi^B ) = 0, \; \forall \; B.
\end{equation}
The only non zero 
element of the YM stress energy tensor is
\begin{equation}
T_{\s}{}^{\r} = \frac{1}{4 \pi}  F_{\s c}^B F^{C \r c} G_{BC} 
 = \frac{-P^4}{4 \pi Q^2} (\p^k \phi^B)(\p_k \phi^C) G_{BC},
\end{equation}
$G_{BC}$ being the invariant metric in $Lie(G)$.
The Lovelock-YM equations therefore reduce to (\ref{egbm2}) added to
\begin{equation}
 \left( \sum_{p=1}^{[(d-1)/2]} \a_p v(p,n) \l^{p-1} \right) \; \; K_{\s}{}^{\r}
 = \frac{-2 P^4}{Q^2} (\p^k \phi^B)(\p_k \phi^C) G_{BC}.
\end{equation}
Inserting (\ref{kk}) this gives
\begin{equation} \label{egbm3}
F_2(\l) \left[ \p^k \p_k H + \frac{\l n(n+2)}{4 P^2} H \right] =
\left( \frac{4 P^{2-n/2}}{ Q }\right)  (\p^k \phi^B) (\p_k \phi^C) G_{BC}
\end{equation}
Eqs (\ref{egbm1}), (\ref{egbm2}) and (\ref{egbm3}) are the 
Lovelock-YM equations for a PFGW.
Notice that, if $G~=~U(1)$,  then we have to drop the gauge indices $A,B,
C,...$ and replace $G_{BC} \to 1$ everywhere, and
(\ref{egbm3}) reduces to  the same {\em form} 
 as equation (23) of \cite{r}, although 
with a different interpretation for the parameters. Some of 
the solutions that will be found in the following sections may, therefore, 
be seen also as containing and generalizing the results found in \cite{r}.
Also, since $F_{ab}F^{ab}=0$ for the electromagnetic field above,
 our electrovac solutions 
 are also solutions of the Lovelock-Born-Infeld field equations \cite{lbi}.\\
As in \cite{r}, it is convenient to
 write the n-dimensional flat space laplacian in
(\ref{egbm3}) in terms of hyperspherical coordinates, $(\xi,\theta^{\alpha})$,
with the radial variable
$\xi$ given by $\xi=\sqrt{z_k z^k} =\sqrt{\delta_{jk}z^j z^k}$, and the $n-1$
angular variables $\theta^{\alpha}$ restricted so that the (n-1)-sphere is covered in the standard way.  We then
have,
\begin{equation} \label{hyper1}
\p^k \p_k H = \xi^{1-n} \p_{\xi}(\xi^{n-1}
\p_{\xi}H)+\xi^{-2} \Delta_{S_{n-1}}H
\end{equation}
where $\Delta_{S_{n-1}}$ is the Laplacian on the (n-1)-hypersphere. This
suggests immediately a separation of variables,
\begin{equation} \label{hyper2}
 H(\sigma,\xi,\theta^{\alpha}) =
\sum_{L,\ell_{n-2},...,\ell_1}\tilde{H}_{L,\ell_{n-2},...,\ell_1}(\sigma,\xi) Y_{L,\ell_{n-2},...,\ell_1}(\theta^{\alpha})
\end{equation}
where $Y_{L,\ell_{n-2},...,\ell_1}(\theta^{\alpha})$ are scalar spherical harmonics on $S_{n-1}$, satisfying (see, e.g., \cite{spharm}),
\begin{equation} \label{hyper3}
\Delta_{S_{n-1}} Y_{L,\ell_{n-2},...,\ell_1}(\theta^{\alpha})=-L(L+n-2) Y_{L,\ell_{n-2},...,\ell_1}(\theta^{\alpha}),
\end{equation}
the integers $L,\ell_{n-2},...,\ell_1$ 
satisfy $L\geq\ell_{n-2}...\geq \ell_2 \geq|\ell_1|$, and $L=0,1,2,...$.
Using (\ref{hyper2}) and (\ref{hyper3}), 
for the {\em vacuum} case (\ref{egbm3}) is reduced to
\begin{equation} \label{hyper4}
\frac{\p^2 \tilde{H}}{\p \xi^2}+\frac{(n-1)}{\xi}\frac{\p\tilde{H}}{\p\xi} +
 \left[\frac{\lambda n(n+2)}{4 (1+\lambda \xi^2/4)^2 }-\frac{L(L+n-2)}{\xi^2}\right] \tilde{H}=0
\end{equation}
where $\tilde{H}$ stands for $\tilde{H}_{L,\ell_{n-2},...,\ell_1}$.

\section{Solutions for $\lambda = 0$} \label{zc}

Equations (\ref{egbm2}) and 
 (\ref{egbm3}) simplify considerably when the curvature (\ref{riemannwf})
of the wave front is zero. 
In this section we exhibit  a number of interesting $\l =0$ solutions.
Notice from (\ref{egbm1})  that $\lambda = 0$ is possible 
only if the cosmological constant $\a_o = 0$. Note also from 
(\ref{egbm2}) and (\ref{egbm3}) that, for $\l=0$, the Lovelock-YM
equations are independent of the $\a_p, p>1$, and thus 
are solutions of the Einstein-YM equations with zero cosmological constant.

\subsection{Vacuum solutions}

 For $\l=0$, the solution of the vacuum equation  (\ref{hyper4}) is
\begin{equation} \label{HLeq0}
\tilde H(\sigma,\xi)= f_1(\sigma)   \xi^L  +f_2(\sigma) \xi^{2-n-L}
\end{equation}
and the general solution of (\ref{egbm3}) is obtained by linear 
combinations with suitable spherical harmonics
\begin{equation} \label{le0vac}
 H(\sigma,\xi,\theta^{\alpha}) =
\sum_{L,\ell_{n-2},...,\ell_1}
\left[ f^{(1)}_{L,\ell_{n-2},...,\ell_1}(\sigma) \;\xi^L + 
f^{(2)}_{L,\ell_{n-2},...,\ell_1}(\sigma) \;\xi^{2-n-L} \right] 
Y_{L,\ell_{n-2},...,\ell_1}(\theta^{\alpha})
\end{equation}
where the $f^{(1,2)}_{L,\ell_{n-2},...,\ell_1}$ are arbitrary functions 
of $\s$.

\subsection{Lovelock-Yang-Mills solutions}
For  $\lambda = 0$  we notice that
(\ref{egbm2}) reduces 
to 
\begin{equation} \label{egbm2p}
\p_k (  \p^k \phi^B ) = 0,
\end{equation}
 and, therefore, 
the general solution for $\phi^B$ may be written,
\begin{equation} \label{psi2}
 \phi^B(\sigma,\xi,\theta^{\alpha}) =
\sum_{L,\ell_{n-2},...,\ell_1}
\left[ f^{(1)B}_{L,\ell_{n-2},...,\ell_1}(\sigma)\; \xi^L + 
f^{(2)B}_{L,\ell_{n-2},...,\ell_1}(\sigma)\; \xi^{2-n-L} \right] 
Y_{L,\ell_{n-2},...,\ell_1}(\theta^{\alpha})
\end{equation}
In this case (\ref{egbm3}) reduces to
\begin{equation} \label{egbm3a}
\p^k \p_k H  =-
\left( \frac{1 }{ \a_1 Q(\s,z^i)}\right)  (\p^k \phi^B) (\p_k \phi^C) G_{BC}
\end{equation}
which is of Poisson type for $H$. However,
since $Q$ depends in general non-trivially on $z^i$, this equation is 
difficult to solve,
unless we impose some restrictions on $\phi$ and $Q$.
We consider first an interesting example of
such restrictions that lead to $pp-$waves.

\subsubsection{Lovelock-Yang-Mills pp-waves}

$pp-$waves arise when $\l=0$ if we further choose $\beta_i(\s)=0$ for all $i$,
and, without loss of generality, set $\alpha(\sigma)=1$.
The metric reduces to (\ref{ppgeo}) and
  (\ref{egbm3}) to
\begin{equation} \label{egbm3app}
\p^k \p_k H  =-
\left( \frac{1  }{ \a_1}\right)  (\p^k \phi^B) (\p_k \phi^C)G_{BC}.
\end{equation}
One can check that,in view of (\ref{egbm2p}), the general solution of
 (\ref{egbm3app}) is 
\begin{equation} \label{psi3pp}
 H(\sigma,\xi,\theta) = H_0(\sigma,\xi,\theta)-
\left(\frac{1}{2 \a_1 }\right) \phi^B (\sigma,\xi,\theta)
\phi^C (\sigma,\xi,\theta) G_{BC},
\end{equation}
with $H_0$ an arbitrary solution of 
$\p_k (  \p^k H_0 ) = 0$. 
As explained above, these 
  Lovelock-YM, $pp-$waves are also Einstein-Yang-Mills $pp-$waves 
and, as such, have been studied before. Non-abelian plane waves in Minkowski
space-time were first studied in \cite{coleman}, whereas 
Einstein-Yang-Mills $pp-$waves appeared in \cite{guven},
and were reconsidered in the supergravity context in \cite{gg2}.
For the Einstein-Maxwell case they were also given in \cite{r}.

\subsubsection{Lovelock-Yang-Mills plane fronted waves}

If we allow $ \beta_i \neq 0$, we may find solutions of (\ref{egbm3a}) for $\lambda=0$,
that generalize those found by Obukhov \cite{r}. Recalling that for $\lambda=0$ we have $Q=\alpha(\sigma)+\sum {\beta_i(\sigma) z^i}$, we look for solutions for $\phi^B$, such that,
\begin{equation} \label{confi1}
\partial_k \phi^B = F^B_k(\sigma,Q).
\end{equation}
Namely, such that $\partial_k \phi^B$ depends on $z^i$ only through $Q$. Since we also require $\partial^k \partial_k \phi^B = 0$, we must have,
\begin{equation} \label{confi2}
 F^B_k(\sigma,Q) = \sum_{\ell m}{\epsilon_{k \ell m}(\sigma)\frac{\partial {\cal{F}}^B_{\ell}(\sigma,Q)}{\partial z^m}}
\end{equation}
where ${\cal{F}}^B_n(Q)$ are arbitrary function of $\sigma,Q$, and $\epsilon_{k\ell m}$
is totally antisymmetric in all
its indices, but, otherwise, arbitrarily dependent on $\sigma$. We therefore have,
\begin{equation} \label{confi3}
 \partial_k \phi^A \partial_k \phi^B G_{AB}  =
 \epsilon_{k\ell m}\epsilon_{kij} \beta_m \beta_j {\cal{\widetilde{F}}}^A_{\ell}{\cal{\widetilde{F}}}^B_i G_{AB}
\end{equation}
where a sum over all repeated indices is implied, and
${\cal{\widetilde{F}}}^A_n=\partial {\cal{F}}^A_n(\sigma,Q)/\partial Q$.

With this ansatz, the right hand side of (\ref{egbm3a}) is, essentially, an arbitrary function of $Q$, since $\sigma$ may be taken as constant, as far as solving (\ref{egbm3a}) for $H$ is concerned. If we set $H(\sigma,z^k)= H(\sigma,Q)$, (\ref{egbm3a}) takes the form,
\begin{equation} \label{egbm3b}
\beta_k \beta^k\frac{\partial^2 H}{\partial Q^2}  = \frac{{\cal{S}}(\sigma,Q)}{Q}
\end{equation}
where ${\cal{S}}(\sigma,Q)$ is obtained by replacement of (\ref{confi3}) in (\ref{egbm3a}). This equation can then be solved, in principle, by quadratures in $Q$. As an example, consider the case where the functions ${\cal{F}}^A_i(\sigma,Q)$ are polynomials in $Q$. This implies that ${\cal{S}}(\sigma,Q)$ is also a polynomial in $Q$ of a certain degree $N$. If we write,
\begin{equation} \label{source1}
{\cal{S}}(\sigma,Q)= \sum_{k=0}^N {{\cal{C}}_k(\sigma) Q^k},
\end{equation}
we find that a particular solution of (\ref{egbm3b}) is given by,
\begin{equation} \label{part1}
H = \frac{1}{\beta_j \beta^j}
\left[
{\cal{C}}_0 Q (\ln{Q} -1)+
\sum_{k=1}^N {{\cal{C}}_k  \frac{Q^{(k+1)}}{k(k+1)}} \right]
\end{equation}

The general solution is then obtained adding to (\ref{part1}) the homogeneous solutions for $H$  (\ref{HLeq0},\ref{le0vac}). We notice also that if we restrict the gauge group to $U(1)$ (electromagnetism), and consider only the case $N=0$ in (\ref{source1}), the solution (\ref{part1}) takes the same form as that given in \cite{r} for the analogous Einstein-Maxwell case.

\section{Solutions for $\lambda \neq 0$} \label{nzc}

In this Section we consider PFGWs for which the curvature of the
wave fronts is non-zero. The pure gravity solutions are worked out
in full detail. For Lovelock-Yang-Mills we find the general solution for the YM fields
satisfying the source free condition, and the restricted case of ``hyperspherical symmetry" is briefly considered.

\subsection{Vacuum solutions}

The general solution of the vacuum equation 
(\ref{hyper4}) for $\lambda\neq 0$ may be written in the form,
\begin{equation} \label{hyper5}
\tilde{H}(\sigma,\xi)  = f_1(\sigma) H_1^L(\xi)+f_2(\sigma) H_2^L(\xi),
\end{equation}
where 
\begin{eqnarray} \label{hyper6a}
H_1^L(\xi) & = &  \xi^L {}_2F_1\left(\frac{n}{2}+1,-\frac{n}{2};\frac{n}{2}+L;\frac{\lambda \xi^2}{(4+\lambda\xi^2)}\right)   \\
\label{hyper6b}
H_2^L(\xi)& = &  \xi^L \left[\frac{(4+\lambda\xi^2)^{n/2+1}}{(4-\lambda\xi^2)^{n+L}}\right] {}_2F_1\left(\frac{1+n+L}{2},\frac{n+L}{2};\frac{n+3}{2};\frac{(4+\lambda\xi^2)^2}{(4-\lambda\xi^2)^2}\right).
\end{eqnarray}
The general solution of (\ref{egbm3}) is then,
\begin{equation} \label{ln0vac}
 H(\sigma,\xi,\theta^{\alpha}) =
\sum_{L,\ell_{n-2},...,\ell_1}
\left[ f^{(1)}_{L,\ell_{n-2},...,\ell_1}(\sigma) \;H_1^L(\xi) + 
f^{(2)}_{L,\ell_{n-2},...,\ell_1}(\sigma) \;H_2^L(\xi) \right] 
Y_{L,\ell_{n-2},...,\ell_1}(\theta^{\alpha}),
\end{equation}
where the $f$'s are arbitrary.
We notice that $H_1^L$ is regular for $\xi=0$, while $H_2^L$ 
is singular. For $L=0$, and $L=1$ we have, respectively,
\begin{equation} \label{H1Leq0}
H_1^{L=0}= f_1(\sigma)  \left(1-\frac{\lambda \xi^2}{4}\right)\left[1+\frac{\lambda \xi^2}{4}\right]^{-n/2}
\end{equation}
\begin{equation} \label{H1Leq1}
H_1^{L=1}= f_1(\sigma)   \xi \left[1+\frac{\lambda \xi^2}{4}\right]^{-n/2}
\end{equation}
Notice that from (\ref{egbm1}) and (\ref{egbm3}), a PFGW {\em vacuum}
 solution of a generic 
Lovelock theory is also a solution of Einstein gravity with a suitable 
cosmological constant, the
 theory treated in \cite{r}. Under this identification,
the particular solutions given as (24) and (25) in \cite{r}
coincide in form with (\ref{ln0vac}) above if we set all $f^{(i)}_{L...}$ to zero for $L>1$
harmonics, and  recall that in general we have $z^k=\xi$
times a linear combination of $L=1$ spherical harmonics.
 Similarly, it can be checked that for $L=0$, the solutions given by $H_2$ 
coincide in form with the solutions given as $H_2$ in \cite{r}.

\subsection{Lovelock-Yang-Mills solutions}
We consider now gravity coupled to a YM field 
 for $\lambda \neq 0$. 
Changing again to hyperspherical coordinates, and separating variables as in
 (\ref{hyper2}), the relevant part of (\ref{egbm2}) takes the form,
\begin{equation} \label{egbm2a}
\frac{\partial^2 \phi^B}{\partial \xi^2}
+\left[\frac{4(n-1)-(n-3) \lambda \xi^2}{\xi (4+\lambda \xi^2)}\right]
\frac{\partial \phi^B}{\partial \xi}
-\frac{L(L+n-2)}{\xi^2} \phi^B = 0.
\end{equation}
 The general solution of this equation, for $L\neq 0$,
may be written in terms of hypergeometric functions. For $n+L$ even we have,
\begin{eqnarray}
\label{Lpneven}
\phi^B (\s ,\xi ) & = & {C_1^B}(\s)
\,\xi ^{L}\,(4 + \lambda \,\xi ^{2})^{(n
 - 1)}\, {}_2F_1 \left( n - 1 + L, \,{\displaystyle \frac {n
}{2}}; \,{\displaystyle \frac {n}{2}}  + L; \, -
{\displaystyle \frac {\lambda \,\xi ^{2}}{4}} \right) \nonumber \\
& & + {C_2^B}(\s)\,
  {}_2F_1 \left(1 -
{\displaystyle \frac {n}{2}}  - {\displaystyle \frac {L}{2}} , \,
{\displaystyle \frac {L}{2}}; \,{\displaystyle \frac {1}{2}}
; \, - {\displaystyle \frac {\left( - 4 + \lambda \,\xi ^{2}\right)^{2}}{16
\,\lambda \,\xi ^{2}}} \right)
\end{eqnarray}
while, for $n+L$ odd, the solution may be written in the form,
\begin{eqnarray}
\label{Lpnodd}
\phi^B (\s ,\xi ) & = & {C_1^B}(\s)
\,\xi ^{L}\,(4 + \lambda \,\xi ^{2})^{(n
 - 1)}\, {}_2F_1 \left( n - 1 + L, \,{\displaystyle \frac {n
}{2}}; \,{\displaystyle \frac {n}{2}}  + L; \, -
{\displaystyle \frac {\lambda \,\xi ^{2}}{4}} \right) \nonumber \\
& & +  {C_2^B}(\s){\displaystyle \frac{1}{\xi}}
 {\left( 1- {\displaystyle \frac {\lambda \,\xi
^{2}}{4}}\right)
\,{}_2F_1 \left({\displaystyle \frac {3}{2}
}  - {\displaystyle \frac {n}{2}}  - {\displaystyle \frac {L}{2}
} , \,{\displaystyle \frac {1}{2}}  + {\displaystyle \frac {L}{2}
}; \,{\displaystyle \frac {3}{2}} ; \, - {\displaystyle
\frac {( - 4 + \lambda \,\xi ^{2})^{2}}{16\,\lambda \,\xi ^{2}}}
\right)}
\end{eqnarray}

We remark that both in (\ref{Lpneven}) and (\ref{Lpnodd}) the hypergeometric functions in the second term in the right hand sides reduce to polynomials of $L$ and $n$ dependent degree in their arguments.

The solution for $L=0$ may be written as
\begin{equation}
\label{Leq0a}
\phi^B (\s ,\xi ) ={C_1^B}(\s)
  +  {C_2^B}(\s){\displaystyle \frac{1}{\xi}}
 {\left( 1- {\displaystyle \frac {\lambda \,\xi
^{2}}{4}}\right)
\,{}_2F_1 \left({\displaystyle \frac {3}{2}
}  - {\displaystyle \frac {n}{2}}   , \,{\displaystyle \frac {1}{2}}  ; \,{\displaystyle \frac {3}{2}} ; \, - {\displaystyle
\frac {( - 4 + \lambda \,\xi ^{2})^{2}}{16\,\lambda \,\xi ^{2}}}
\right)}
\end{equation}
and the same remark as for (\ref{Lpneven}) and (\ref{Lpnodd}) is valid here for odd $n$. We also notice that for odd $n$ (and $L=0$) we may also set,
\begin{equation}
\label{Leq0b}
\phi^B (\s ,\xi ) ={C_1^B}(\s)
  +  {C_2^B}(\s) \xi^{(2-n)}
\,{}_2F_1 \left(1
 - {\displaystyle \frac {n}{2}}   , 2-n  ; \,2-{\displaystyle \frac {n}{2}} ; \, - {\displaystyle
\frac {\lambda \,\xi ^{2} }{4}}
\right)
\end{equation}

The solutions for $L=0$ coincide with those given in \cite{r}.

It is clear that finding general solutions for (\ref{egbm3}), for the general forms of $\phi^B$, $G_{AB}$, and $Q$ may be a difficult task. However, if we restrict to $L=0$ (``hyperspherical symmetry''), then $ \phi^B$ is a function of only $\sigma$, and $\xi$, and from (\ref{egbm2a}) we find,
\begin{equation}
\label{Leq0c}
\frac{\partial \phi^B (\s ,\xi )}{\partial z^k} ={C^B}(\s) \frac{z^k}{\xi^n} P^{n-2}
\end{equation}
where $C^B(\s)$ is an arbitrary function of $\sigma$. Replacing in (\ref{egbm3}), we find,
\begin{equation} \label{Leq0d}
\p^k \p_k H + \frac{\l n(n+2)}{4 P^2} H = {\cal{A}}  \frac{P^{3n/2-2}}{\xi^{2n-2} Q}
\end{equation}
where ${\cal{A}}$ depends only on $\sigma$, (and the other parameters of the theory), and is determined from (\ref{egbm3}) once (\ref{Leq0c}) is given. Equation (\ref{Leq0d}) is identical in form to the equation that results in the restricted Einstein-Maxwell case, analyzed by Obukhov in \cite{r}. The general solution is given there and will not be repeated here.

\section{Degenerate Lovelock theories}

A given Lovelock theory is characterized by the set of coefficients 
$\a_p$ in (\ref{lovelockt}), or, equivalently, by the polynomial $F_1(\l)$ 
in (\ref{egb2})-(\ref{egb2_2}). {\em Degenerate} Lovelock
 theories are those for which $F_1(\l)$ has one or more 
(real) roots with multiplicity
greater than one. As already noticed in \cite{w2,w3}, 
the Lovelock equations do not fix the dynamics entirely if the theory is
degenerate. Here we would like to comment briefly on PFGWs in degenerate 
Lovelock theories.\\
From (\ref{kk}) and (\ref{egb2}), a
 possible vacuum solution of Lovelock's equations
is $H=0$, $\l$ a root
of $F_1(\l)$. In this case one can see that $K_{ab}{}^{cd}=0$, then 
 $R_{ab}{}^{cd} = \lambda (\d_a{}^c \d_b{}^d - \d_b{}^c \d_a{}^d)$
(see (\ref{riemann1})-(\ref{riemann2})). This is an 
 $(n+2)$ dimensional homogeneous space-time, and thus 
 locally isometric
to (A)dS or Minkowski space-time, depending on the sign  of $\l$. 
Homogeneous vacuum solutions of Lovelock gravity 
are well known. They were first obtained for Einstein-Gauss-Bonnet gravity 
(Lovelock theory with $\a_p=0$ for all $p>2$) 
in \cite{w1,bd}, generalized in \cite{w2}, 
and reconsidered, e.g, in \cite{w3}.\\
Now suppose the theory is degenerate, and let $\l_d$ be a 
{\em double} root of $F_1$. Note from (\ref{egb2_2}) that 
 $F_2(\l) = (-2/(n(n+1))
F_1{}'(\l)$, then  $F_1(\l_d)
=F_2(\l_d)=0$, and  ${\cal{G}}_b{}^a = 0$ for {\em any} $H$. 
 These vacuum solutions contain
$H$ as an extra arbitrary function. This is the degeneracy
noticed in \cite{w1,w2,w3}.
Note also that no PFGWs solutions with $\l=\l_d$ can be obtained 
if we add a YM field. 
If  $\l=\l_s$ is a single root of $F_1(\l)$, 
Lovelock-YM PFGWs with wave fronts of curvature $\l_s$ do exist in this case,
and we recover  the usual degrees of freedom -both for pure 
gravity and Lovelock-YM-, as  (\ref{egbm3}) is 
a nontrivial equation for $H$. Since non-degenerate 
theories have no double roots, 
they always give a non trivial equation for $H$.\\
 A highly degenerate Lovelock
theory was considered in \cite{atz,btz}, for which $f_1(\l) 
\propto (\l-\Lambda_{ADS})^{[(d-1)/2]}, \Lambda_{ADS} < 0$. 
The only homogeneous solution in this 
case is AdS. Other interesting solutions are the 
asymptotically AdS black holes, known as BTZ black holes.
 One important feature of BTZ theories 
is that the action 
is locally invariant under the AdS group, enlarging the 
usual local Lorentz symmetry of gravity theories. Since $F_1$
does not have single roots in a BTZ theory, 
PFGWs cannot be constructed if we couple 
a YM field to a BTZ theory.

\section{Comments and conclusions}

In this paper we have given prescriptions for the
construction of  plane fronted gravitational waves 
in Lovelock-Yang-Mills theory with arbitrary Lovelock coefficients.
These are $n+2$ dimensional space-times with a shear, expansion and twist free 
null congruence, perpendicular to wave fronts of constant curvature $\l$. 
In higher dimensional Einstein gravity with a non vanishing  cosmological constant 
$\L$, these waves always  exist, and  $\l=\L$. 
In Lovelock's  theory, on the other hand,  PFGWs exist only
if the polynomial $F_1$ introduced in equation (\ref{egb2}) has real 
roots, each real root being an allowed value for the  curvature 
of the wave front. As is well known,  a homogeneous vacuum 
solution of Lovelock's equations 
with curvature $\l$ exist for each 
real root of $\l$ of $F_1$ \cite{bd,atz}. We have shown in this paper 
that  a PFGW propagating in 
this homogeneous space-time  is always possible. As an example, consider 
Einstein-Gauss-Bonnet theory ($ \a_2 \neq 0,
 \a_p=0$, for all $p>2$ in (\ref{lovelockt})).
The two possible values of $\l$ are 
\begin{equation} \label{lgb}
\l_{\pm} = {\frac {-n \left( n+1 \right) {  \alpha_1} \pm \sqrt{ n^2 \left( n+1 \right)^2  {  \alpha_1}^{2}-4 \left( n+1
 \right) n \left( n-1 \right)  \left( n-2 \right) {  \alpha_2}{  
\alpha_0}}}
{ 4 \left( n+1 \right) n \left( n-1 \right)  \left( n-2 \right) {  
\alpha_2}}}
\end{equation}
This means that there are no solutions if 
$$ \a_o \a_2 > \frac{n(n+1)}{4 (n-1) (n-2)} {\a_1}^2 $$
In the limit $\a_2 \to 0$ (small string tension), (\ref{lgb}) we have, 
$$
\l_{\pm} = -\frac{(1 \pm 1) \a_1}{4(n-2)(n-1)} {\a_2}^{-1} \pm  \frac{\a_o}{2n(n+1) \a_1}
+ {\cal{O}} (\a_2).
$$
Therefore, $\l_-$ approaches $\l$ of  Einstein's theory PFGWs, whereas $\l_+$ 
becomes unbounded.

\section*{Acknowledgments}
We are grateful to Gary Gibbons  and Ricardo Troncoso
 for useful comments on a preliminary version of this paper.
This work was supported in part by grants of the National University of
C\'ordoba and Agencia C\'ordoba Ciencia (Argentina). It was also supported in
part by grant NSF-INT-0204937 of the National Science Foundation of the US. The
authors are supported by CONICET (Argentina).


\begin{thebibliography}{99}
\bibitem{Lovelock} D. Lovelock, {\it Jour. Math. Phys.} {\bf 12} (1971) 498.
\bibitem{str1} D. J. Gross and E. Witten, Nucl. Phys. {\bf B277}, 1 (1986);
B. Zumino, Phys. Rep. {\bf 137}, 109 (1986);
B. Zwiebach, Phys. Lett. {\bf 156B}, 915 (1985);
D. Friedan, {\it Phys. Rev. Lett.} {\bf 45} (1980) 1057;
I. Jack, D.Jones and N. Mohammedi, Nucl. Phys. {\bf B322}, 431 (1989);
C. Callan, D. Friedan, E. Martinec and M. Perry, Nucl. Phys. {\bf B262}, 
593 (1985).
\bibitem{str2} Y. Cai and C. Nu\~nez, Nucl. Phys. {\bf B287}, 279 (1987);
D.J. Gross and J.H. Sloan, Nucl. Phys. {\bf B291}, 41 (1987); 
A.A. Tseytlin,   
 Nucl.Phys.{\bf B584} 233,(2000),  
hep-th/0005072;
M.T. Grisaru, A.E. van de Ven and D. Zanon;  Phys. Lett. {\bf B173}, 423 (1986);   
 Nucl. Phys. {\bf B277}, 409 (1986); Phys. Lett. {\bf B177}, 347 (1986); M.D.
 Freeman, C.N. Pope, M.F. Sohnius and K.S. Stelle,  
 Phys. Lett. {\bf B178}, 199 (1986); Q. Park and D. Zanon,  
Phys. Rev. {\bf D35}, 4038 (1987). 
\bibitem{m} T. Damour, hep-th/0504153. 
\bibitem{jeno} S. Cnockaert and  Marc Henneaux, hep-th/0504169. 
\bibitem{malda} D. Berenstein, J. Maldacena and  H. Nastase, AIP
 Conf.Proc.646:3-14,2002.
\bibitem{r}  Y. N. Obukhov, Phys.Rev.{\bf D69}024013 (2004), gr-qc/031012.
\bibitem{joro} G. T. Horowitz and A. R. Steif, 
{\it Phys. Rev. Lett.} {\bf 64} (1990) 260.
\bibitem{gg1} G.W. Gibbons and P.J. Ruback,  {\it Phys. Lett.} {\bf B171} (1986)
 390.
\bibitem{guven} R. G\"uven,  Phys.Rev. {\bf
 D19} 471 (1979).
\bibitem{gg2} M. Cariglia, G.W. Gibbons, R. Guven and C.N. Pope  Class.Quant.Grav.21:2849-2858,2004, hep-th/0312256.
\bibitem{g}  M. Aiello, R. Ferraro and  G. Giribet, Phys.Rev. {\bf
 D70}:104014,2004, gr-qc/0408078.
\bibitem{cm} Rong-Gen Cai, Da-Wei Pang, Anzhong Wang
Phys.Rev. {\bf D70} (2004) 124034.
\bibitem{pph} A. Coley, R. Milson, N. Pelavas, V. Pravda, A. Pravdova and R. Zalaletdinov, Phys.Rev. {\bf D67}:104020,2003,
gr-qc/0212063. 
\bibitem{orr} I.Ozsvath, I.Robinson and K.Rozga, J. Math. Phys. {\bf 26} 1755 
(1985).
\bibitem{gdp}
 A. Garc\'{\i}a D\'{\i}az and J.F. Plebanski, J. Math. Phys. {\bf 22}, 2655
(1981).
\bibitem{k1} W. Kundt, Z. Phys. {\bf 163}, 77 (1961); 
H. Stephani, D. Kramer, M. Maccallum, C. Hoenselaers and 
E. Herlet, {\em Exact solutions to Einstein's field equations} Second edition, 
Cambridge University Press (2003), chapter 31.
\bibitem{k2} Ji\v{r}\'{i} Bi\v{c}ak and Ji\v{r}\'{\i} Podolsk\'y, 
J. Math. Phys. {\bf 40}, 4495 (1999).
\bibitem{lbi} M. Born and L. Infeld, Proc. Roy. Soc. Lond {\bf A144}, 425 (1934).
\bibitem{spharm}
 A. Higuchi,{\it Jour. Math. Phys.} {\bf 28} (1987) 1553 (Erratum
 ibid.43:6385,2002).
\bibitem{coleman} S. Coleman, {\it Phys. Lett.} {\bf B70} (1977)
 59.
\bibitem{w2} J. T. Wheeler, Nucl.Phys.{\bf B273} 732 (1986)
\bibitem{w3} B. Whitt, Phys. Rev. {\bf D38} 3000, (1988).
\bibitem{w1} J. T. Wheeler, Nucl.Phys.{\bf B268} 737,(1986).
\bibitem{bd} D. G. Boulware  and S. Deser   {\it Phys. Rev. Lett.} {\bf 55} (1985), 2656.
\bibitem{btz} M. Ba\~nados, C. Teitelboim and J. Zanelli  {\it
Phys. Rev.} {\bf D49} (1994), 975.
\bibitem{atz} R. Aros, R. Troncoso  and J. Zanelli  {\it Phys. Rev.} 
 {\bf D63} (2001), 084015.
\end{thebibliography}
\end{document}